\documentclass[fleqn,twoside]{article}
\usepackage{espcrc2}
\usepackage{graphicx}
\usepackage{epsfig}

\newcommand{\beq}{\begin{equation}}
\newcommand{\eeq}{\end{equation}}
\newcommand{\bea}{\begin{eqnarray}}
\newcommand{\eea}{\end{eqnarray}}

\newcommand{\AmS}{{\protect\the\textfont2
  A\kern-.1667em\lower.5ex\hbox{M}\kern-.125emS}}

\title{Elsevier instructions for the preparation of a
       2-column format camera-ready paper in \LaTeX}

\title{Physics Beyond the Standard Model (Theory): \\
        Introducing the Little Higgs}
\author{Martin Schmaltz\address{Physics Department, Boston University,
                                Boston MA 02215, USA \\
                                email: schmaltz@bu.edu \\
\vskip-1.86in Plenary talk given at ICHEP 2002 - Amsterdam, The Netherlands -
        Jul 29 2002 - BUHEP-02-35 \vskip1.8in
}
                       \thanks{Work supported by the Department of Energy, 
                               Contract DE-FG02-91ER-40676} }

\begin{document}


\begin{abstract}
Little Higgs theories are an exciting new possibility for
physics at TeV energies. In the Standard Model the Higgs
mass suffers from an instability under radiative corrections.
This ``hierarchy problem'' motivates much of current physics
beyond the Standard Model research. Little Higgs
theories offer a new and very promising solution to this problem
in which the Higgs is naturally light as a result of
non-linearly realized symmetries. This article reviews some of
the underlying ideas and gives a pedagogical introduction
to the Little Higgs. The examples provided are taken from the
paper "A Little Higgs from a Simple Group",
by D.E. Kaplan and M. Schmaltz \cite{KS}.

\vspace{1pc}
\end{abstract}

\maketitle

\section{Introduction}

Given the Standard Model's remarkable success in accurately describing
Physics at length scales ranging from atomic scales all the way down to
the shortest currently probed scales of about $10^{-18}$ m it may appear
puzzling that so much theoretical and experimental work is devoted to
discovering physics beyond the Standard Model (SM)

In the first part of
this talk I review the basic argument - the hierarchy problem -
which motivates much of Physics Beyond the SM (PBSM) research.
I then argue that the hierarchy problem can be used to make
specific predictions with regards to the quantum numbers of
new particles we might expect to discover at future colliders,
and I present those predictions.
In the second part of my talk, I give a pedagogical introduction to a
recently discovered new solution to the hierarchy problem, the
``Little Higgs'' mechanism.

\section{The hierarchy problem of the Standard Model}

The predictions of the Standard Model have been probed directly
at LEP and at the Tevatron up to energy scales of order of
a few hundred GeV. Lower bounds on masses of new particles BSM
are approximately \cite{PDG}
\bea
&100& {\rm\!\!\! GeV\ \ new\ leptons} \nonumber \\
&200&  {\rm\!\!\! GeV\ \ new\ quarks}  \nonumber\\
&300&  {\rm\!\!\! GeV\ \ leptoquarks} \nonumber \\ 
&700& {\rm\!\!\! GeV\ \ }  W' {\rm\ or\ } Z' \nonumber
\eea
In addition, precision measurements at LEP, SLC and
the Tevatron have indirectly probed scales between 1 and 10 TeV
and found no significant disagreement with SM predictions.
Finally, bounds on new
flavor and CP violating couplings indirectly probe scales
as high as 100 to 1000 TeV for new flavor violating couplings
or order one at those scales.

Altogether, the experimental evidence shows
that the SM works extremely well, and one might be tempted to postulate
a ``minimal scenario'': we will discover the SM Higgs boson
with a mass somewhere between the current lower bound of 114 GeV
\cite{PDG} and the triviality bound near 500 GeV,
and there is no new PBSM up to very high scales. In particular,
this would imply that the LHC will not see any PBSM. In the following I
argue that this pessimistic option is not only disfavored
by the data but is also extremely unlikely because it implies
a very delicate and unnatural fine-tuning of parameters.

The bounds from precision data require that new physics at the
TeV scale is approximately flavor preserving and sufficiently
weakly coupled so as to not generate large radiative corrections.
In particular, new particles with masses of order a TeV are not very
constrained. Most radiative corrections to precision observables from
weakly coupled TeV scale particles are suppressed by factors of
$(M_W/TeV)^2$ relative to loops with SM particles. Experiments are
sensitive at the level of SM loops but not at the level of small
corrections to them.
Well-known examples for such new physics which may be hiding at
the TeV scale are supersymmetry, vector-like extra generations,
new gauge bosons or extended Higgs sectors.

Nevertheless, one might argue that the simplest possibility is to
assume that the SM remains valid even beyond scales at which
it has been directly tested. What is wrong with that possibility,
why are we convinced that new physics will be discovered at the
TeV scale? The answer is provided by the hierarchy problem which
I briefly review here; not because the hierarchy problem is unfamiliar,
but because I would like to formulate the problem in a way
which best motivates the Little Higgs solution. In addition,
I want to connect the hierarchy problem to future experiments.

At the LHC the 1-10 TeV energy scale will be probed directly for the
first time. Thus an important question to answer is whether it is
natural for the SM to be valid up to these scales (or if we can
expect to discover new physics at the LHC). To make the argument, 
let us assume that the SM is valid up to a cut-off scale of
$\Lambda=10$ TeV. At even higher energies new physics takes over,
which implies that we do not know how to compute loop diagrams with 
momenta larger than $\Lambda$, thus we will cut such loops off at
$\Lambda$. The hierarchy problem arises from the fact that there
are quadratically divergent loop contributions to the Higgs
mass which drive the Higgs mass to unacceptably large values unless
the tree level mass parameter is finely tuned to cancel the large
quantum corrections.

The most significant of these divergences come from three sources.
They are - in order of decreasing magnitude - one-loop diagrams
(Figure 1.) involving the top quark, the $SU(2)\times U(1)$ gauge
bosons, and the Higgs itself.

\begin{figure}[htb]
\centerline{\epsfysize=2.0in
{\epsffile{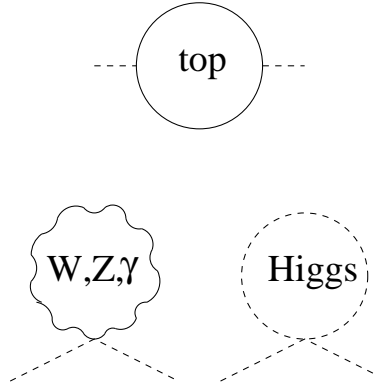}}}
\caption[]{\it The most significant quadratically divergent contributions 
to the Higgs mass in the Standard Model. }
\label{fig:smloops}
\end{figure}

All other
quadratically divergent diagrams involve small coupling constants
and do not significantly contribute for $\Lambda =$ 10 TeV.
The contributions from the three diagrams are
\beq
-{3 \over 8 \pi^2} \lambda_t^2 \Lambda^2\ \sim\ -(2\, TeV)^2
\eeq
from the top loop,
\beq
{1 \over 16 \pi^2} g^2 \Lambda^2\ \sim \ (700\, GeV)^2
\eeq
from the gauge loop, and
\beq
{1 \over 16 \pi^2} \lambda^2 \Lambda^2\ \sim \ (500\, GeV)^2
\eeq
from the Higgs loop. Thus the total Higgs mass is approximately
\beq
m_h^2=m_{tree}^2 - [ 100 - 10 - 5]\ (200 GeV)^2 \ .
\eeq

In order for this to add up to a Higgs mass of order a few hundred
GeV as required in the SM fine tuning of one part in 100
(see Figure 2.) is required. This is the hierarchy problem.

\begin{figure}[htb]
\centerline{\epsfysize=2.0in
{\epsffile{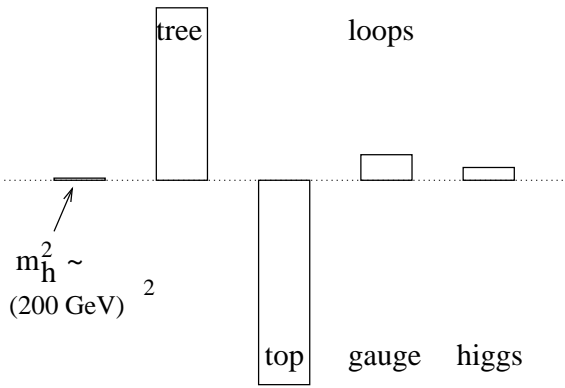}}}
\caption[]{\it The fine tuning required to obtain an acceptable
Higgs mass in the Standard Model with cut-off 10 TeV. }
\label{fig:hierarchy}
\end{figure}

Is the SM model already fine tuned given that we have experimentally
probed to near 1 TeV and found no PBSM? Setting $\Lambda=$ 1 TeV in
the above formulas we find that the most dangerous contribution
from the top loop is only about (200 GeV)$^2$. Thus no fine tuning is required,
the SM with no new physics up to 1 TeV is perfectly natural, and
we should not be surprised that we have not yet seen deviations from
it at colliders.

In the following, we will turn the argument around and use the hierarchy 
problem to predict what forms of new physics exist at what scale
in order to solve the hierarchy problem. Consider for example
the correction to the Higgs mass form the top loop. Limiting this
contribution to be no larger than about 10 times the Higgs mass
(limiting fine-tuning to less than 1 part in 10) we find a maximum
cut-off for $\Lambda= $ 2 TeV. In other words, we predict the existence
of new particles with masses less than or equal to 2 TeV which cancel
the quadratically divergent Higgs mass contribution from the top quark.
In order for this cancellation to occur naturally, the new particles 
must be related to the top quark by a symmetry. In practice this
means that the new particles have to have similar quantum numbers
to top quarks. Thus the hierarchy problem predicts a new multiplet
of particles with mass below 2 TeV which carry color and are
easily produced at the LHC. In supersymmetric extensions
these new particles are of course the top squarks.

The contributions from gauge loops also need to be canceled
by new particles which are related to the SM $SU(2) \times U(1)$
gauge bosons by a symmetry. The masses of these states are
predicted to be at or below 5 TeV for the cancellation to be natural.
Similarly, the Higgs loop requires new states related to the Higgs
at 10 TeV. We see that the hierarchy problem can be
used to obtain specific predictions.

\bea
{\rm SM\ loop} \quad &{\rm maximum\ mass\ of\ new\ particles}
\nonumber \\
{\rm top} \quad &{\rm 2\ TeV} \nonumber \\
{\rm weak\ bosons} \quad &{\rm 5\ TeV} \nonumber \\
{\rm Higgs} \quad &{\rm 10\ TeV}\nonumber 
\eea

\subsection{Supersymmetry and the hierarchy}

One successful approach to solving
the hierarchy problem is based on supersymmetry (SUSY).
Loosely speaking, in SUSY every quadratically
divergent loop diagram in Figure 1. has a superpartner, a diagram
with superpartners running in the loop (Figure 3.). The
diagrams with superpartners exactly cancel the quadratic divergences
of the SM diagrams. Generically, this happens because
superpartner coupling constants are related to SM coupling 
constants by supersymmetry, but superpartner loops have the opposite
sign from their SM partner because of opposite spin-statistics.

\begin{figure}[htb]
\centerline{\epsfysize=2.4in
{\epsffile{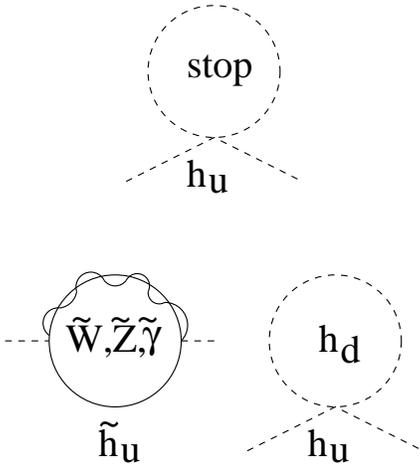}}}
\caption[]{\it Superpartner diagrams which cancel quadratic divergences
of the Standard Model. }
\label{fig:susyloops}
\end{figure}

In the limit of unbroken supersymmetry the diagrams cancel completely.
If weak scale SUSY occurs in nature superpartner masses
softly break the supersymmetry. Then the cancellation only
takes place above the scale of superpartner masses $M_{SUSY}$.
Below $M_{SUSY}$ only the SM particles exist, thus there the
SM loop diagrams (Figure 1) are not canceled but the cut-off
$\Lambda$ is replaced by $M_{SUSY}$.

Experimental bounds on superpartner masses are somewhere in the
few hundred GeV range, much lower than the upper bound of 2 TeV
from fine tuning constraints top loops. Nevertheless,
the MSSM is already somewhat fine tuned.
The problem arises from the experimental lower bound on the Higgs
mass. As is well known, the tree level Higgs mass in the MSSM is
bounded from above by the Z mass. Radiative corrections from stop loops
give a positive contribution and can lift the Higgs mass above
the experimental bound of 114 GeV. However, a large enough correction
requires heavy stops which in turn reintroduce fine tuning.

Current experimental bounds force a minimum
fine tuning on the order of $10\%$.

\subsection{SUSY, the only solution to the hierarchy problem?}

Until recently it was widely believed that supersymmetry represents
the only possible weakly coupled solution to the hierarchy
problem. This belief was based on a lack of known alternatives, bolstered
by a ``folk theorem''. The ``folk theorem'' loosely states that
supersymmetry is the only theory in which quadratic divergences
cancel without tuning. The ``folk proof'' roughly goes as follows:
{\it i.} boson and fermion loops have opposite signs due to a
minus sign in the Feynman rules for fermion loops --
{\it ii.} therefore cancellation of divergences only occurs between
boson and fermion loops --
{\it iii.} in order for the cancellation to be natural the boson
and fermion loops need to be related by a symmetry: supersymmetry.

However, the folk theorem is wrong. Amusingly, a
counter example to step {\it ii.} in the above ``proof'' occurs
in the MSSM itself. Because it is instructive
let us briefly look at this example. The MSSM extends the
Higgs sector of the SM to a two Higgs doublet model.
The tree level quartic couplings for the Higgses arise from integrating
out the D-auxiliary fields in the $SU(2)\times U(1)$ gauge vector
multiplets. Looking for example only at the contribution from hypercharge,
the relevant terms in the Lagrangian are
\beq
{\cal L} = \frac12 D^2 + \frac{g}2 ( h_u^* D h_u - h_d^* D h_d) \ 
\label{eqn:dterm}
\eeq
giving the usual quartic D-term in the Higgs potential
\beq
V = \frac12 \left( \frac{g}2 \right)^2 ( h_u^* h_u - h_d^* h_d)^2
\eeq
The cancellation of quadratic divergences is most easily understood
be keeping the D-auxiliary fields in the theory. Then, the ``quartic
coupling'' contained in eq.~(\ref{eqn:dterm}) leads to three
distinct diagrams (Figure 4.).

\begin{figure}[htb]
\centerline{\epsfysize=2.4in
{\epsffile{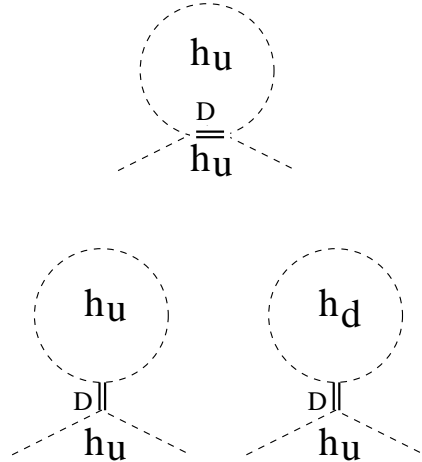}}}
\caption[]{\it Higgs loops in the MSSM. }
\label{fig:mssmloops}
\end{figure}

The first cancels against fermionic
loops with gauginos as predicted by the folk theorem. However, the
other two diagrams cancel between each other. The cancellation occurs
between diagrams with only bosons. It requires no fine tuning because the
diagrams are proportional to the hypercharges of the Higgses in
the loop which are opposite in sign and equal in magnitude (by gauge
invariance and supersymmetry). Thus we see that even in the MSSM some
of the quadratic divergences cancel between bosons, with the required
difference in sign simply arising from a difference in signs between
coupling constants.

\section{Introducing the Little Higgs}

In this part of the talk I give a brief pedagogical introduction
to Little Higgs (LH) theories. The material covered is almost entirely
contained in the recent literature which I briefly review here:
Little Higgs theories are realizations of an old idea to
stabilize the Higgs mass by making the Higgs a pseudo-Goldstone boson
resulting from a spontaneously broken approximate symmetry.
Early attempts at
constructing such models \cite{georgipais,DGK} were not entirely
successful and quadratic divergences to the Higgs mass remained.
The first successful model which canceled all relevant quadratic
divergences based on the pseudo-Goldstone idea
was constructed by Arkani-Hamed, Cohen and Georgi in \cite{ACG}.
Subsequently, simpler and more elegant models were constructed.
The models may be described by their global symmetry breaking patterns,
$SU(5)/SO(5)$ \cite{ACKN2}, $SU(6)/SP(6)$ \cite{LSS}, the
minimal moose $SU(3)^2/SU(3)$ \cite{ACKN}, and general mooses
$SU(3)^n/SU(3)^k$ \cite{GW}.
Model building constraints
on specific UV completions of LH have been discussed in
\cite{lane,CES}, and preliminary studies of LH phenomenology
appeared in \cite{ACGW}. More recently, a new variant of the
LH mechanism was discovered which does not require duplication
of gauge groups \cite{KS}.
At this conference LH were already presented
in two talks by Lane \cite{Lanetalk} and Wacker \cite{Wackertalk}.

Let us begin by discussing LH theories from a phenomenological
point of view. It should be clear from the previous section that
they necessarily involve new particles related to the top quark,
the $SU(2)\times U(1)$ gauge bosons and the Higgs. In LH theories
the masses of these new states are given by a single scale
$f$ at which global symmetries are spontaneously broken. The
spectrum of a generic theory is summarized in Figure 5.
\begin{figure}[htb]
\centerline{\epsfysize=1.8in
{\epsffile{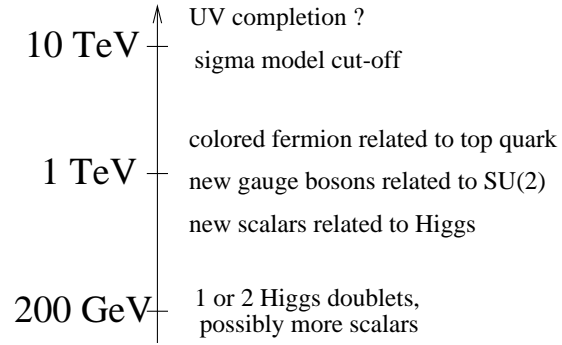}}}
\caption[]{\it Generic Little Higgs Spectrum. }
\label{fig:LHspectrum}
\end{figure}
From a purely mechanical point of view the cancellation of quadratic
divergences in LH models is very simple. As in the case of supersymmetry
there are diagrams with new particles in loops which precisely cancel
the quadratic divergences. Consider for example the loop involving
top quarks. LH Higgs theories contain a new vector like quark
$\chi$ with mass of order $f$ and couplings which are
related to the top Yukawa coupling $\lambda_t$. The relevant
interactions of this new fermion are
\beq
\lambda_t f \left(1-\frac12 {h^*h \over f^2} \right)
\chi_R^\dagger \chi_L\ +\ h.c.
\eeq 
They allow a new quadratically divergent contribution to
the Higgs mass from Figure 6 which involves the same integral
as the top loop and is proportional to the coupling constants
\beq
\lambda_t^* f \ (-\frac{\lambda_t}{2f}) = -\frac12|\lambda_t^2| \ .
\eeq
There are two such diagrams which together
exactly cancel the top loop divergence
\begin{figure}[htb]
\centerline{\epsfysize=2.4in
{\epsffile{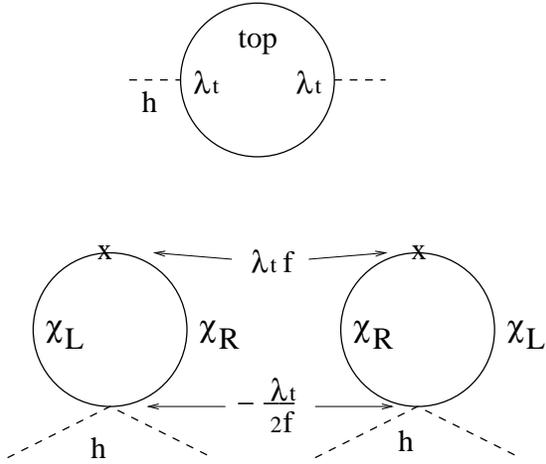}}}
\caption[]{\it Canceling the top quark loop. }
\label{fig:LHcancelation}
\end{figure}

As advertised, a fermion loop cancels a fermion loop.
Similarly, the gauge and Higgs loops are canceled by diagrams
with new bosons in loops.
\begin{figure}[htb]
\centerline{\epsfysize=2.4in
{\epsffile{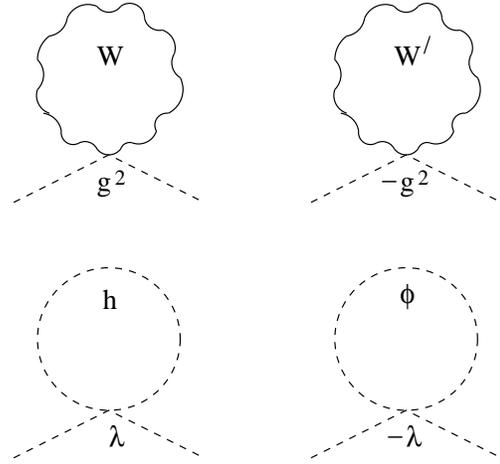}}}
\caption[]{\it Canceling the gauge and Higgs loops. }
\end{figure}

Note that this alone is of course no solution to the hierarchy
problem. Without a symmetry which enforces the relations between
coupling constants required for the cancellations
one has only shifted the fine tuning from the Higgs mass to the
coupling constants of the new fields. However, Little Higgs 
theories have such a symmetry built in. To discover it we will
take a little detour.

\subsection{Higgs as a Pseudo-Goldstone boson}

In this section we identify the symmetry which makes the
identification of coupling constants of the previous discussion
radiatively stable and therefore ``technically natural''.
To do so we first briefly review Goldstone bosons.

Massless Goldstone bosons always arise when global symmetries are
spontaneously broken by the vacuum. Consider for example the
theory of a complex scalar field $\Phi$ with
potential ${\cal V}={\cal V}(|\Phi|)$.
This potential preserves a global $U(1)$ symmetry
$\Phi \rightarrow e^{i \epsilon} \Phi$. If the potential
induces a vacuum expectation value for $<\Phi>=f$
then the $U(1)$ symmetry is spontaneously broken and a massless
Goldstone boson arises. To see this explicitly, it is convenient to
parameterize $\Phi= (v+r)e^{i \theta/f}$, where $r$ and $\theta$ are
real fields. Because of the global $U(1)$ symmetry $\theta$ can be removed
from the potential by a space-time dependent $U(1)$ transformation
with $\epsilon = - \theta/f$. The resulting Lagrangian does not
contain $\theta$ except in derivative interactions. Therefore
$\theta$ has no potential and in particular also no mass.
The ``radial mode'' $r$ does obtain a mass from the potential
and can be integrated out (Figure 7.).
\begin{figure}[htb]
\centerline{\epsfysize=1.6in
{\epsffile{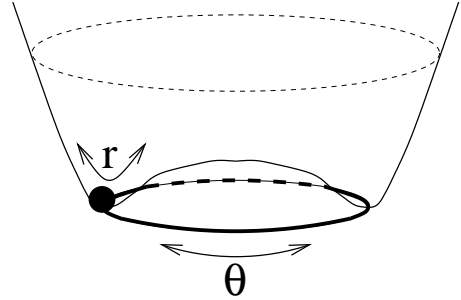}}}
\caption[]{\it The ``Mexican hat'' potential for $\Phi$. The
black dot represents the vacuum expectation value $f$, $r$ is the radial
mode and $\theta$ the Goldstone boson. }
\label{fig:mexhat}
\end{figure}
Note that this argument is based only on the existence of the
$U(1)$ symmetry and is therefore stable under radiative corrections.

To summarize, we now have a simple mechanism for generating massless
scalar fields. Unfortunately, the symmetry needed forbids all 
non-derivative couplings. Thus $\theta$ is not a good toy example
for a light Higgs, it has no quartic, gauge or Yukawa couplings.

Furthermore, $\theta$ also has the wrong quantum numbers to be
the SM Higgs. This is easily fixed by generalizing to
non-abelian symmetry breaking. Consider for example the
breaking of a global $SU(3)$ symmetry to $SU(2)$ by an
expectation value for a complex triplet field
$<\Phi^T>=\ <\left(\phi_1,  \phi_2, \phi_3 \right)>
= \left( 0, 0, f \right)$. In this vacuum $\Phi$ is
conveniently parameterized as
\beq
\Phi=e^{i \theta/f} \left( \begin{array}{l}
0  \\ 0 \\ f \end{array} \right)
\eeq
where $\theta=\theta^a\ T^a$ is a hermitian $3\times 3$ matrix containing the
properly normalized five Goldstone bosons from the breaking
of $SU(3)\rightarrow SU(2)$
\beq
\theta={1 \over \sqrt{2}}
\left( \begin{array}{cc} 
\!\!\begin{array}{ll} 0 & 0 \\ 0 & 0 \end{array} 
& \!\!h \\ h^\dagger & \!\!0 \end{array} \right)
+{\eta \over 4} 
\left( \begin{array}{rrr} 1&0 &0 \\0 &1&0\\ 0& 0&\!\!\!-2 \end{array}
\right)  .
\eeq
Under the unbroken $SU(2)$ the four real fields in $h$ transform as
a complex doublet whereas $\eta$ is a singlet. As in the abelian
case, by performing a space-time dependent $SU(3)$ rotation it is
possible to remove all non-derivative interactions of the
Goldstone fields. This is still very different from the SM Higgs boson,
our ``Higgs'' $h$ is an $SU(2)$ doublet but it has no gauge,
quartic, or Yukawa interactions and is exactly massless.

To add these couplings we need to explicitly break the
global $SU(3)$ symmetry. Explicit symmetry breaking introduces
non-derivative couplings of the Goldstone bosons, they become
pseudo-Goldstone bosons. If the explicit symmetry breaking stems
only from a small ``spurion'' parameter $\epsilon$ then all non-derivative
couplings are proportional to the spurion. This remains true even
including radiative corrections. Thus breaking of global symmetries
by small spurions gives us control over quantum corrections.
In the diagram of Figure 7. a small breaking of the symmetry
corresponds to a small tilt of the ``Mexican hat'' proportional
to $\epsilon$. 

This is all very nice and allows us to control the mass of
pseudo-Goldstone bosons, however -- unfortunately -- the SM Higgs
couplings are not small. Thus even though radiative corrections
are proportional to these couplings, this does not sufficiently
suppress the contributions to the Higgs mass;
i.e. simply adding the SM Higgs couplings and thereby explicitly
breaking the $SU(3)$ symmetry leads to the radiative corrections
of Figure 1. which are too large $\sim \lambda_t^2/16\pi^2\  \Lambda^2$.

The new model building idea which led to the construction of Little
Higgs models is collective breaking of symmetries. Instead of breaking
the symmetry with a single coupling, one introduces two couplings in such
a way that each coupling by itself preserves sufficient amount of symmetry
to guarantee the masslessness of the pseudo-Goldstone boson.
Schematically, we add two new sets of interactions ${\cal L}_i$
to the $SU(3)$ preserving Lagrangian ${\cal L}_0$
\beq
{\cal L}= {\cal L}_0 + \epsilon_1 {\cal L}_1 + \epsilon_2 {\cal L}_2\ ,
\eeq
where each term is chosen such that by itself it preserves an
$SU(3)$ symmetry but that together they break $SU(3)$ explicitly.
Therefore radiative corrections to the Higgs mass are necessarily
proportional to both spurions $\epsilon_1$ and $\epsilon_2$. In the
example we will study below, the fact that both spurions are required
implies that quadratically divergent contributions arise only at two loops
$\sim \epsilon_1^2/16\pi^2\ \epsilon_1^2/16\pi^2\ \Lambda^2$
which is sufficiently small for $\Lambda=$ 10 TeV even for
$\epsilon_i \sim 1$.

\subsection{Toy Little Higgs theory}

In order to study the LH mechanism without the notational
complexity require by the many fields of the SM let us
imagine a toy world without hypercharge in which the
only fermions are top and bottom quarks with their normal
Yukawa couplings and $SU(3)\times SU(2)$ gauge interactions.
Particle theorists of this toy world have constructed a
``Toy Standard Model'' which
suffers from the same hierarchy problem as the real SM.
In order to protect the toy SM Higgs from quadratic divergences
one can introduce an $SU(3)$ symmetry which is spontaneously broken
to $SU(2)$ at the scale $f\sim$ 1 TeV.
The Higgs consists of four of the resulting Goldstone
bosons as described before.

We generate Yukawa and gauge couplings without reintroducing
quadratic divergences by using collective symmetry breaking.
This is achieved by introducing two $SU(3)$ groups which are
each spontaneously broken to $SU(2)$ by expectation values for
two scalar triplets $\Phi_1$ and $\Phi_2$.

Concretely, $SU(2)$-weak is enlarged into an $SU(3)$ gauge group.
Two scalar fields $\Phi_i$ are both charged under this $SU(3)$
and both obtain expectation values. In order to understand the
collective breaking, imagine turning off the gauge coupling to
either one of the triplets. In this limit, there are two $SU(3)$
symmetries, one gauged, one global. Thus the model
has two $SU(3)$ symmetries which are explicitly broken to the
diagonal gauge group by the gauge couplings. But note that the
gauge couplings to both $\Phi_1$ and $\Phi_2$ are required
for the breaking. The $SU(3)$ gauge couplings of $\Phi_1$
and $\Phi_2$ play the role of the spurions $\epsilon_1$ and $\epsilon_2$.

It is convenient to parameterize the light fields in the $\Phi_i$ as
\bea
\Phi_1=e^{i \theta/f}
\left( \begin{array}{l}
0  \\ 0 \\ f \end{array} \right) \nonumber \\
\Phi_2=e^{-i \theta/ f}
\left( \begin{array}{l}
0  \\ 0 \\ f \end{array} \right) \ .
\label{eq:phis}
\eea
Here $\theta$ are the pseudo-Goldstone bosons corresponding to
breaking of the ``axial'' $SU(3)$. The eaten Goldstone bosons
corresponding to the diagonal ``vector'' $SU(3)$ have been
removed by a gauge transformation (to unitary gauge). The expectation
values $f$ of the $\Phi$'s have been chosen to be equal for simplicity.
Note that neither of the $SU(3)$'s here have anything to do
with color.

Ignoring the singlet $\eta$, the pseudo-Goldstones are
\beq
\theta=\frac{1}{\sqrt{2}}
\left( \begin{array}{cc} 
\!\!\begin{array}{ll} 0 & 0 \\ 0 & 0 \end{array} 
& \!\!h \\ h^\dagger & \!\!0 \end{array} \right) \ ,
\eeq
i.e. a complex Higgs doublet which is charged under the
unbroken $SU(2)$ gauge group. Explicitly, the $SU(2)$
gauge interactions for $h$ stem from the $SU(3)$
gauge couplings of the $\Phi$'s
\bea
 \left[(\partial_\mu+i g A_\mu)\Phi_1\right]^\dagger
 (\partial_\mu+i g A_\mu)\Phi_1 \nonumber \\
\ +\  \left[(\partial_\mu+i g A_\mu)\Phi_2\right]^\dagger
 (\partial_\mu+i g A_\mu)\Phi_2 \ .
\eea
In calculating the quadratic divergence from the gauge
interactions it is most convenient to work in a manifestly
$SU(3)$ invariant formalism. Then quadratic divergences simply
come from a diagram with external $\Phi$'s and an
$SU(3)$ gauge boson loop (the first diagram in Figure 8.).
\begin{figure}[htb]
\centerline{\epsfysize=1.4in
{\epsffile{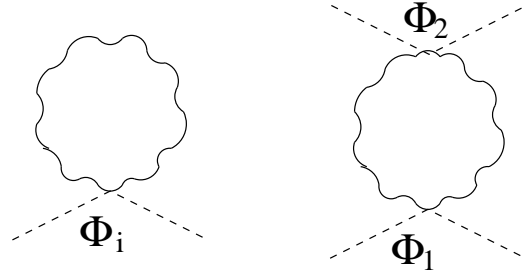}}}
\caption[]{\it Quadratically and logarithmically divergent diagrams
from the $SU(3)$ gauge interactions. The first diagram does not
produce a Higgs mass.}
\label{fig:su3gauge}
\end{figure}
The quadratic divergence of this
diagram produces the operators
\beq
{g^2\over 16 \pi^2} \Lambda^2
\left(\Phi_1^\dagger\Phi_1+\Phi_2^\dagger\Phi_2 \right)
\eeq
which respects both $SU(3)$ symmetries and therefore does
not contain couplings of the pseudo-Goldstone bosons.
This can also be seen explicitly by substituting eq.~(\ref{eq:phis}).
The logarithmically divergent second diagram in Figure 8.
gives rise to an operator which explicitly breaks
$[SU(3)]^2 \rightarrow SU(3)$ and therefore contains a mass
for $h$
\bea
{g^4\over 16 \pi^2} \left| \Phi_1^\dagger \Phi_2 \right|^2
{\rm log}(\Lambda^2/f^2) \rightarrow \nonumber \\
{g^4 f^2 \over 16 \pi^2}\, {\rm log}(\Lambda^2/f^2) \ h^\dagger h\ ,
\label{eq:gaugelog}
\eea
which is of order 100 GeV for $f\sim$ 1 TeV.

The mechanism for adding a top Yukawa coupling without introducing
a quadratic divergence relies on the same principle. Since the
weak interactions are embedded in an $SU(3)$ gauge group we need
to introduce left-handed quarks in an $SU(3)$ triplet
$(t_L,b_L,\chi_L)$. The first two components are the usual quark
doublet whereas the third component is a new quark.
In addition to the triplet we also introduce three right
handed quark singlets, two correspond to the right
handed top and bottom quarks and one is the
right handed component $\chi_R$ of the vector like heavy quark $\chi$
which is responsible for canceling the top loop divergence.
To produce a mass for $\chi$ and a Yukawa coupling for the
top quark we add 
\beq
\frac{\lambda_t}{\sqrt{2}}\ ( \chi_1 \Phi_1^\dagger +  \chi_2 \Phi_2^\dagger) 
\left( \begin{array}{l}
t_L  \\ b_L \\ \chi_L \end{array} \right) 
\label{eq:yuk}
\eeq
where -- again for simplicity -- we have taken the two Yukawa
couplings equal and factored out a $\sqrt{2}$ in order to simplify
subsequent equations. Expanding the $\Phi_i$ to second order in
the Higgs field and defining the two linear combinations
\beq
\chi_R=\frac{1}{\sqrt{2}}\ (\chi_1+\chi_2) , \quad
t_R=\frac{1}{\sqrt{2}}\ (\chi_1-\chi_2)  ,
\eeq
eq.~(\ref{eq:yuk}) becomes
\beq
\lambda_t f\ (1-{h^\dagger h \over 2 f^2})\ \chi_R^\dagger \chi_L
+ \lambda_t \ t_R^\dagger\, h^\dagger
\left( \begin{array}{l} t_L  \\ b_L \end{array} \right) \ .
\eeq
The first term yields a TeV scale mass $m_\chi=\lambda_t f$ and
a higher-dimensional coupling to the Higgs for the vector-like
quark. The second term is the top Yukawa coupling. Note that
these couplings are exactly what we needed for the cancellation of
the quadratic divergences in Figure 6.
The cancellation of the divergence is again most transparent in
an $SU(3)$ symmetric calculation (Figure 9.).
\begin{figure}[htb]
\centerline{\epsfysize=1.0in
{\epsffile{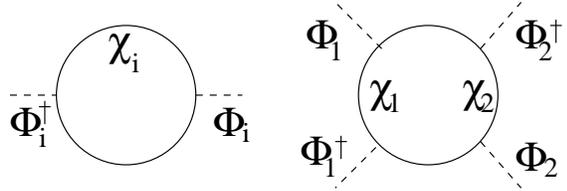}}}
\caption[]{\it $SU(3)$-symmetrized top loops. }
\label{fig:topphi}
\end{figure}
The first diagram is quadratically divergent but
preserves both $SU(3)$'s. Therefore the resulting operator
\beq
{\lambda_t^2\over 16 \pi^2} \Lambda^2
\left(\Phi_1^\dagger\Phi_1+\Phi_2^\dagger\Phi_2 \right)
\eeq
is independent of the pseudo-Goldstones. The second
diagram is only logarithmically divergent and gives a
a contribution
\beq
-{\lambda_t^4\over 16 \pi^2} \left| \Phi_1^\dagger \Phi_2 \right|^2
{\rm log}(\Lambda^2/f^2) \ .
\eeq
This breaks the $SU(3)$ symmetry which protects the Higgs mass, but it's
coefficient is sufficiently small (compare to eq.~(\ref{eq:gaugelog})).

This concludes our discussion of the Little Higgs in the toy
world. To promote the toy LH model to a fully realistic model
requires three extra ingredients
{\it i.} hypercharge, 
{\it ii.} the other quarks and leptons,
{\it iii.} a quartic coupling for the Higgs.
Adding hypercharge is very easy, one simply gauges one of the global
$U(1)$ symmetries of the model. The mechanism for canceling
quadratic divergences is unchanged by this. Adding the remaining
fermions is also straightforward. Note that every SM fermion doublet
has to be promoted to a triplet because of the $SU(3)$ gauge 
symmetry. Adding the Higgs quartic coupling is non-trivial
and requires some additional structure,
which the interested reader can find in
\cite{KS}.

\section{Signatures and Conclusions}

Little Higgs theories predict new states at the TeV scale
which can be seen at the LHC.

Specifically, one generically
finds a {\bf vector-like quark} of charge $2/3$ (up-type) which
is required to cancel the divergence from the top loop.
The new quark can be pair-produced at Hadron colliders as
long as it's mass is within the kinematical reach ($\sim$ 2 TeV).
$\chi$'s are expected to decay predominantly to $h+t$, $W+b$
and $Z+t$. In addition we expect {\bf new gauge bosons} which cancel
the $SU(2)\times U(1)$ gauge loops. Their quantum numbers appear
to be model dependent. Generically they carry weak charges
and may be electrically charged. They couple with weak coupling
strength and can be singly produced at colliders. They have
couplings to light fermions and can mix with the SM $W$ and $Z$.
These couplings typically lead to the strongest constraints
on LH models from electroweak precision tests.
Finally, one also expects {\bf new scalars} with masses somewhere 
in the 100 GeV - 2 TeV range. The quantum numbers of the scalars 
are model dependent but additional Higgs doublets are often found.
The new scalars couple most strongly to the Higgs,
to the weak gauge bosons and to third generation fermions.
In some models one of the neutral scalars is stable because of
an unbroken discrete symmetry providing
an excellent cold dark matter candidate.

In conclusion, I reiterate that despite the spectacular
success of the SM in describing all particle physics experiments 
to date we have good reasons to believe that new physics will
be discovered in the foreseeable future at the Tevatron or the
LHC. This optimism is based on the belief that the hierarchy problem
requires a TeV scale solution. Until recently, the only known
such solution which is weakly coupled at the TeV scale and therefore
does not suffer from fine tuning problems associated with
precision data was supersymmetry. Thanks to recent developments
supersymmetry has acquired a competitor, the Little Higgs. Many
open questions remain: Are there even simpler Little Higgs models?
Which signatures are generic? Are there compelling UV completions
to Little Higgs models? What is the origin of flavor? What are
the constraints from precision electro-weak fits? ...
Obviously much more theoretical work is needed to answer these
questions, but in the end experiments will have to tell tell us if the
Little Higgs has it's place in nature. 

\section{Acknowledgments}
It is my pleasure to thank N. Arkani-Hamed, R.S. Chivukula,
A.G. Cohen, H. Georgi, D.E. Kaplan, A.E. Nelson and W. Skiba
for introducing me to the Little Higgs. I would also like to
thank the organizers of ICHEP2002 for a perfectly organized
conference and for inviting me to speak. 
This work is supported by the Department of Energy with an OJI
grant and under contract DE-FG02-91ER-40676.

\end{document}